\documentclass[conference]{IEEEtran}
\usepackage{cite}
\usepackage{amsmath,amssymb,amsfonts}
\usepackage{algorithmic}
\usepackage{graphicx}
\usepackage{textcomp}
\usepackage{wrapfig}
\usepackage{subcaption}
\usepackage{xcolor}
\def\BibTeX{{\rm B\kern-.05em{\sc i\kern-.025em b}\kern-.08em
    T\kern-.1667em\lower.7ex\hbox{E}\kern-.125emX}}
\begin{document}

\title{Sketch2Cloth: Sketch-based 3D Garment Generation with Unsigned Distance Fields}

\author{\IEEEauthorblockN{Yi He, Haoran Xie and Kazunori Miyata}
\IEEEauthorblockA{\textit{Japan Advanced Institute of Science and Technology} \\
Ishikawa, Japan}}

\maketitle

\begin{abstract}
3D model reconstruction from a single image has achieved great progress with the recent deep generative models. However, the conventional reconstruction approaches with template mesh deformation and implicit fields have difficulty in reconstructing non-watertight 3D mesh models, such as garments. In contrast to image-based modeling, the sketch-based approach can help users generate 3D models to meet the design intentions from hand-drawn sketches. In this study, we propose Sketch2Cloth, a sketch-based 3D garment generation system using the unsigned distance fields from the user's sketch input. Sketch2Cloth first estimates the unsigned distance function of the target 3D model from the sketch input, and extracts the mesh from the estimated field with Marching Cubes. We also provide the model editing function to modify the generated mesh. We verified the proposed Sketch2Cloth with quantitative evaluations on garment generation and editing with a state-of-the-art approach.
\end{abstract}

\begin{IEEEkeywords}
3D model reconstruction, sketch-based generation, unsigned distance function, garment generation and editing
\end{IEEEkeywords}

\section{Introduction}

With the rapid development of graphics and virtual reality technologies (e.g., metaverse), the modeling of garments and clothes is widely adopted in visual applications such as virtual try-on, online shopping, and 3D character modeling. However, the design processes of garments by professional designers are complex and usually take heavy time costs to design and validate the design outcomes. It is challenging and difficult to design the desired garments for common users without modeling skills and design experience. On the other hand, the users can depict the desired 3D objects using hand-drawn sketches which consist of strokes and subjective abstractions of the target objects. Therefore, sketches are commonly used in the early stage of the design process. For common users, sketch-based approaches can facilitate efficient design activities at a low time cost. In this work, we aim to provide sketch-based model generation and editing for 3D garments.

\begin{figure}[t]
    \centering
    \includegraphics[width=\linewidth]{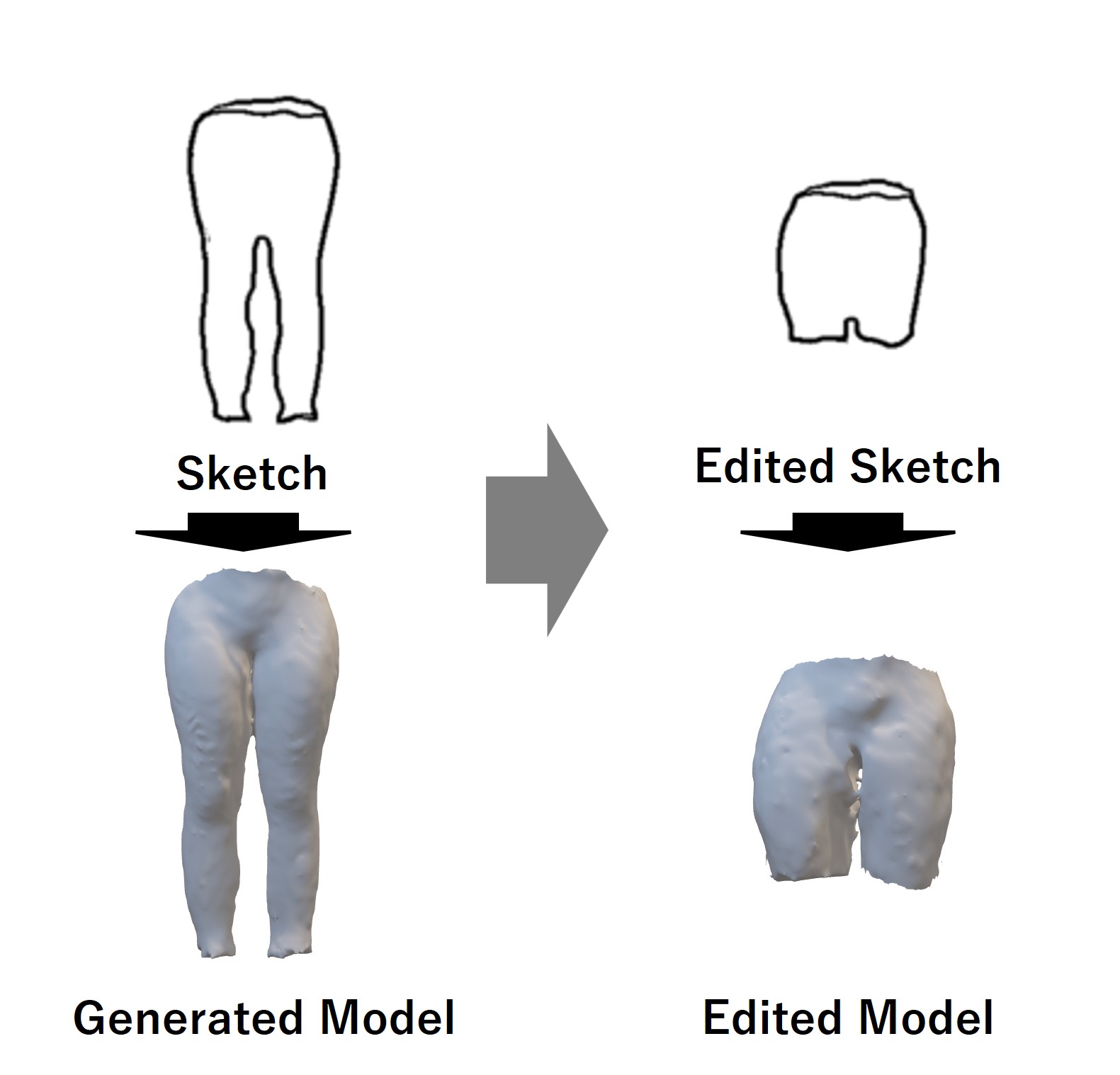}
    \caption{The proposed Sketch2Cloth for sketch-based 3D garment model generation and editing. The proposed system adopts a sketch as input and generates the 3D garment model as output. Sketch2Cloth allows users to edit the generated garment models.}
    \label{fig:overview}
\end{figure}

3D model generation from users' sketches has been extensively studied recently. The representative approaches can map the sketch strokes into 3D space and create the 3D model from the sketch~\cite{Igarashi1999TeddyAS}.  The previous work used sketches to retrieve the target shape from 3D shape dataset by comparing the input sketch with the contour data of 3D models~\cite{wang2015sketch}. Although these previous approaches are considered to be effective in 3D modeling, the acquired shapes may be limited in the existing dataset, and 3D modeling skills are required from  expert users, which makes it infeasible to amateur users.

Recently, deep learning-based approaches can achieve easier and more accurate model generation from a single simple sketch~\cite{zhang2021sketch2model, Guillard2021Sketch2MeshRA}. Sketch2Model tried to generate 3D models from the sketches by deforming a template mesh to a target shape~\cite{zhang2021sketch2model}, which is a common approach for 3D model reconstruction from single-view images.
Skecth2Mesh~\cite{Guillard2021Sketch2MeshRA} utilized the implicit function learning and generated the target shapes by learning a Signed Distance Function (SDF). However, Sketch2Model~\cite{zhang2021sketch2model} may be limited by the topological structure of the template model, and Sketch2Mesh~\cite{Guillard2021Sketch2MeshRA} only for watertight 3D meshes. In this work, we aim to generate non-watertight 3D models from user sketches such as clothing.  To solve this issue, we adopt the implicit field of Unsigned Distance Function (UDF)~\cite{Chibane2020NeuralUD} which is an optimal representation of 3D objects with non-watertight meshes in complex surfaces and scenes.

In this work, we propose Sketch2Cloth to reconstruct 3D garment models from hand-drawn sketches using UDF implicit fields as illustrated in Figure \ref{fig:overview}. The proposed Sketch2Cloth framework utilizes the autoencoder structure to learn UDFs implicit representations. The output of the autoencoder is UDF data, and the mesh can be extracted from the UDF using Marching Cubes method. We also provide a user interface for garment modeling and editing that allows the common users to draw sketches and edit the generated 3D garment models. We achieve the model editing with an encoder of the optimized latent vectors when the users edit the contours of the generated model. Finally, we verified the effectiveness of the proposed Sketch2Cloth system in 3D garment model generation from freehand-drawn sketches through evaluation experiments.

\section{Related Work}

\subsection{Image-based Garment Reconstruction}

With the explosive development of deep learning approaches, numerous approaches for reconstructing 3D models from a single-view image have been explored in recent years, ~\cite{Pavlakos2019ExpressiveBC, Lhner2018DeepWrinklesAA, Alldieck2019Tex2ShapeDF, Zhao2021LearningAU}. The majority of them are template-based methods, where the core of the method is to deform a pre-prepared template mesh into the target shape. In these methods, the results are limited to the topological structure of the template mesh and can only generate rough shapes. Therefore, to represent complex shapes such as clothing, other additional information must be provided. For example, Pavlakos et al.~\cite {Pavlakos2019ExpressiveBC} use a parametric 3D model to generate a human body model by optimizing the parameters of the model to fit 2D features extracted from the image. DeepWrinkles~\cite{Lhner2018DeepWrinklesAA} combines global shape deformation with surface detail by adding fine garment wrinkles to the normal map of a prepared template mesh of clothing. Tex2shape~\cite{Alldieck2019Tex2ShapeDF} considered the shape representation problem as an image style transfer problem and predicted model surface information such as normal maps and added it to the prepared human body models to generate models of clothed people. Although we aimed to provide more complex surface information and learn complex surfaces, these methods cannot break through the limitations of the topological structure of the prepared template mesh.

Apart from the template-based methods, PIFu~\cite{Saito2019PIFuPI} proposed a deep learning framework for clothed body 3D reconstruction based on implicit function learning. In order to retain more surface shape features, all point (x, y, z) information and 2D image pixels in 3D space were locally aligned to successfully reproduce more complex surface structures.
Zhao et al.~\cite{Zhao2021LearningAU} proposed a deep learning framework based on implicit learning-based methods that predicts a set of key points called Anchor Points, which represent features on a fine surface located around the surface, from color images to obtain a more accurate 3D surface representation. These methods are based on learning implicit functions and are not limited by template-based methods. However, these methods use garment photographs as input, which makes sketch-based generation difficult.

\subsection{3D Model Reconstruction with Implicit Functions}

Various 3D reconstruction methods based on implicit learning have been published. Such methods are based on the representation of 3D objects by the implicit field which has the same advantages as point clouds. Typical examples are the Occupancy Field~\cite{Mescheder2019OccupancyNL} and SDF\cite{Park2019DeepSDFLC, MeshSDF} and UDF\cite{Chibane2020NeuralUD, Zhao2021LearningAU}.
These methods can learn continuous shape representations, and the reconstructed 3D models have no resolution limitations. Implicit learning with Occupancy Field and SDF can only generate watertight models, while UDF methods can generate non-watertight models or complex surfaces, including watertight models. Implicit model representation has advantages over template-based methods in that it is not limited by the topological structure of the template, can represent models of any size, and can generate non-watertight models (using UDF). Since garment data is basically non-watertight, it is not possible to determine whether a garment is inside or outside, making implicit function learning with UDFs the best method for this purpose. Guillard et al.~\cite{Guillard2021MeshUDFFA} proposed a method for fast mesh generation from UDFs, which enables efficient mesh extraction from UDFs. In this work, we propose 3D modeling generation for non-watertight clothing meshes using an implicit function learning method with UDF fields.

\begin{figure*}[ht]
    \centering
    \includegraphics[width=\linewidth]{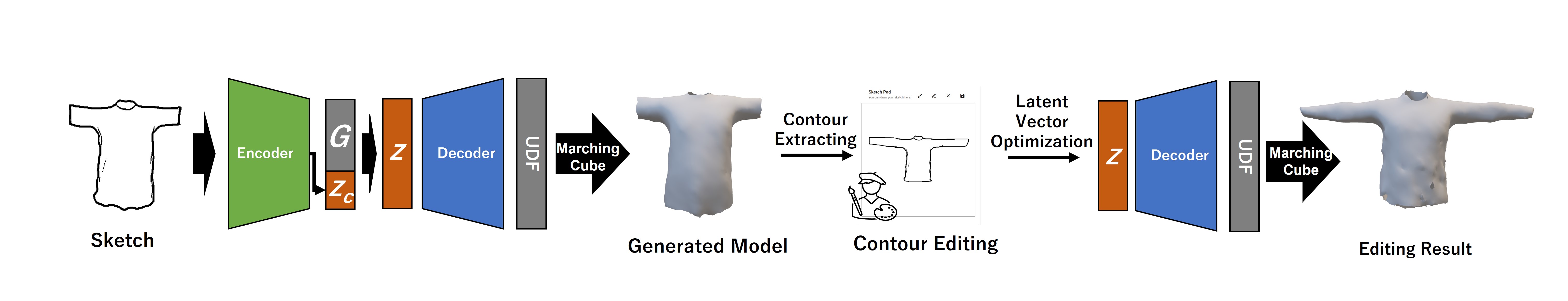}
    \caption{The framework of the proposed Sketch2Cloth system. The proposed system consists of two networks: an encoder that encodes the sketch image into a latent code $z_c$ and a decoder that generates a UDF from the latent vector $z$. The decoder predicts the unsigned distance $d_i^u$ for all $p_i \in G$ based on $z_c$ and outputs the UDF. Then, we use the Marching Cubes method to obtain a 3D model mesh. The user can also edit the generated model with the proposed user interface. }
    \label{fig:framework}
\end{figure*}

\subsection{Sketch-based Content Design}

For both expert and amateur users, sketching is a common and easy way to create content to meet users' design intentions. Recently, the sketch-based approach has been applied in various fields of content creation support, such as image generation~\cite{peng2023difffacesketch,luo2021}, 3D model reconstruction~\cite{Guillard2021Sketch2MeshRA}, and animation controlsF~\cite{sketch2vf19,dualsmoke22,dualmotion22}. For example, Sketch2VF~\cite{sketch2vf19} introduced an interactive user interface for flow design and generated fluid simulations by training a conditional GAN to estimate the velocity field. Similar to this, He et al.~\cite{he2021sketch}  used user sketches to generate normal maps of the target 3D object directly. Furthermore, DualSmoke~\cite{dualsmoke22} adopted a two-stage structure with conditional GAN for Lagrangian coherent structure generation from hand-drawn sketches to help users design the smoke effect without domain knowledge. Peng et al.~\cite{dualmotion22} proposed the motion retrieval system with hand-drawn sketches, while DualFace\cite{dualface22} proposed a portrait drawing interface for general users to draw better portrait paintings. All these applications show us the potential of freehand sketch-based content creation.

Shape generation using sketches has also been widely studied in the computer graphics community. Igarashi et al.~\cite{Igarashi1999TeddyAS} mapped hand-drawn line segments to a specific shape, making it possible to create a 3D model from sketches alone. The explosive development of deep learning has also led to the proposal of various learning-based methods~\cite{Delanoy20183DSU, Lun20173DSR}. These methods usually require information from multiple viewpoints. Zhang et al.~\cite{zhang2021sketch2model} generated a mesh model directly from a sketch, but since it focused on solving the ill-pose problem of the generated geometry (e.g., the direction from which the input sketch was viewed was ambiguous), it did not address complex geometry generation and real-time model generation and editing. Guillard et al.~\cite{Guillard2021Sketch2MeshRA} proposed high-quality mesh model generation and editing using SDF.
However, these methods can only generate watertight models and cannot handle the generation of non-watertight models such as clothing.
Wang et al.~\cite{Wang2018LearningAS} can generate a clothed human body model by learning the latent space shared by the sketch, mesh data, and parameters, and mapping the garment shape to the human body model. Chen et al.~\cite{Chen20213DSG} proposed a method that takes 2D garment sketches as input and combines them with human shape parameters to generate a 3D garment mesh that fits a specific human shape. These approaches were proposed for the generation of a clothed human body that cannot handle only clothing. They also require users to have expertise in 3D modeling and design. To solve these issues, this research aims to develop an easy-to-use user interface that allows users to create sketches and interact with the generated models.

\section{Sketch2Cloth}

The proposed sketch-based garment generation framework, Sketch2Cloth utilizes the autoencoder structure~\cite{MeshSDF} neural network to reconstruct the 3D garment model from users' freehand-drawn sketches. An overview of the framework to generate and edit a 3D model from the input sketch of this research is shown in Figure \ref{fig:framework}. The objective of this study is to reconstruct a 3D model from an input sketch by means of isosurface extraction and triangulated mesh generation for an unsigned distance field (UDF). The implicit UDF field was sampled from the garment model to construct the learning dataset. We also developed a user interface for drawing sketches and viewing and editing the generated models.

\subsection{Unsigned Distance Fields}

We incorporate unsigned distance fields (UDF)~\cite{Chibane2020NeuralUD} for shape surface representation.
Let the 3D space where the 3D model exists be $\Omega$, for any coordinate $p$, we have the equation (\ref{equ:unsigned_distance_field_1}) for the computation of UDF $\Phi(p)$.

\begin{equation}
    \label{equ:unsigned_distance_field_1}
    \Phi(p) = d(p, \partial\Omega), p \in \Omega 
\end{equation}

Here $\partial\Omega$ denotes the boundary of $\Omega$. The distance $d$ from coordinate $p$ to $\partial\Omega$ can be computed by equation (\ref{equ:unsigned_distance_field_2}).

\begin{equation}
    \label{equ:unsigned_distance_field_2}
    d(p, \partial\Omega) = min_{r \in \partial\Omega}\mid p - r \mid
\end{equation}

As the shape surface can be represented by the zero level-set $UDF = 0$, we can get dense point clouds by simply moving $p$ to a new coordinate $q$ from which the distance to the implicit surface is $d$ along the negative gradient direction of UDF. We describe the process in equation (\ref{equ:unsigned_distance_field_3}). Then, the dense point clouds can be used for mesh reconstruction algorithms such as Marching Cubes~\cite{Guillard2021MeshUDFFA}.

\begin{equation}
    \label{equ:unsigned_distance_field_3}
    q = p - d(p, \partial\Omega) \cdot \nabla_pd(p, \partial\Omega)
\end{equation}

\subsection{Dataset Construction}
\label{sec:dataset}

This work used the Multi-Garment dataset~\cite{bhatnagar2019mgn} and DeepFashion3D~\cite{zhu2020deep} dataset to experiment with generating a non-watertight model for reconstructing 3D models of garments. The Multi-Garment dataset contained 328 garment meshes. The dataset was divided into a training set and a test set, with numbers 300 and 28, respectively. The DeepFashion3D dataset contained 2,075 garment meshes and was divided into a training set and a test set, with numbers 1,867 and 208, respectively. To generate UDFs from sketches, we need the pair data of sketches and UDFs.

\subsubsection{Sampling UDF Data}

First, the mesh data is normalized to a maximum coordinate value of $1$ to make it suitable for training. Then, since sampling is performed from both outside and inside the surface, the scale is adjusted to fit a surface of arbitrary radius slightly smaller than $1$ so that sampling can be performed from outside the surface at the farthest distance from the model center, and sampling is performed on the adjusted mesh. Empirically, the scale is set to $0.8$ in this study.

Then, for each $M_i$ mesh, the number of sampled vertices is set to be $N$ and the distance $d_i^u$ from $N$ 3D vertex coordinates $p_i$ to the $M_i$ surface is calculated. There is no limit to the number of samplings, but considering the data processing speed, $N$ is set to $120,000$ to obtain as many samples as possible in as little time as possible. The sampling data consists of random points on the surface, near the surface, and in the bounding box surrounding the mesh. The sampling was concentrated on the surface of the mesh. Specifically, $48,000$ of the samples are taken from coordinates within $0.05$ of the surface, and $32,000$ from coordinates within $0.3$ of the surface, in our implementation. In addition, $24,000$ are sampled randomly from the $M_i$ surface, and the last $16,000$ are sampled within a bounding box with edge length 2 (range $[-1,1]$).

Note that $d_i^u$ is calculated using KDTree~\cite{kdtree}. No special processing (e.g., water-tightening) is required for the mesh $M_i$.

\subsubsection{Sketch-Mesh Paired Data}
\label{sec:sketch_dataset}

\begin{figure}[h]
    \centering
    \includegraphics[width=0.9\linewidth]{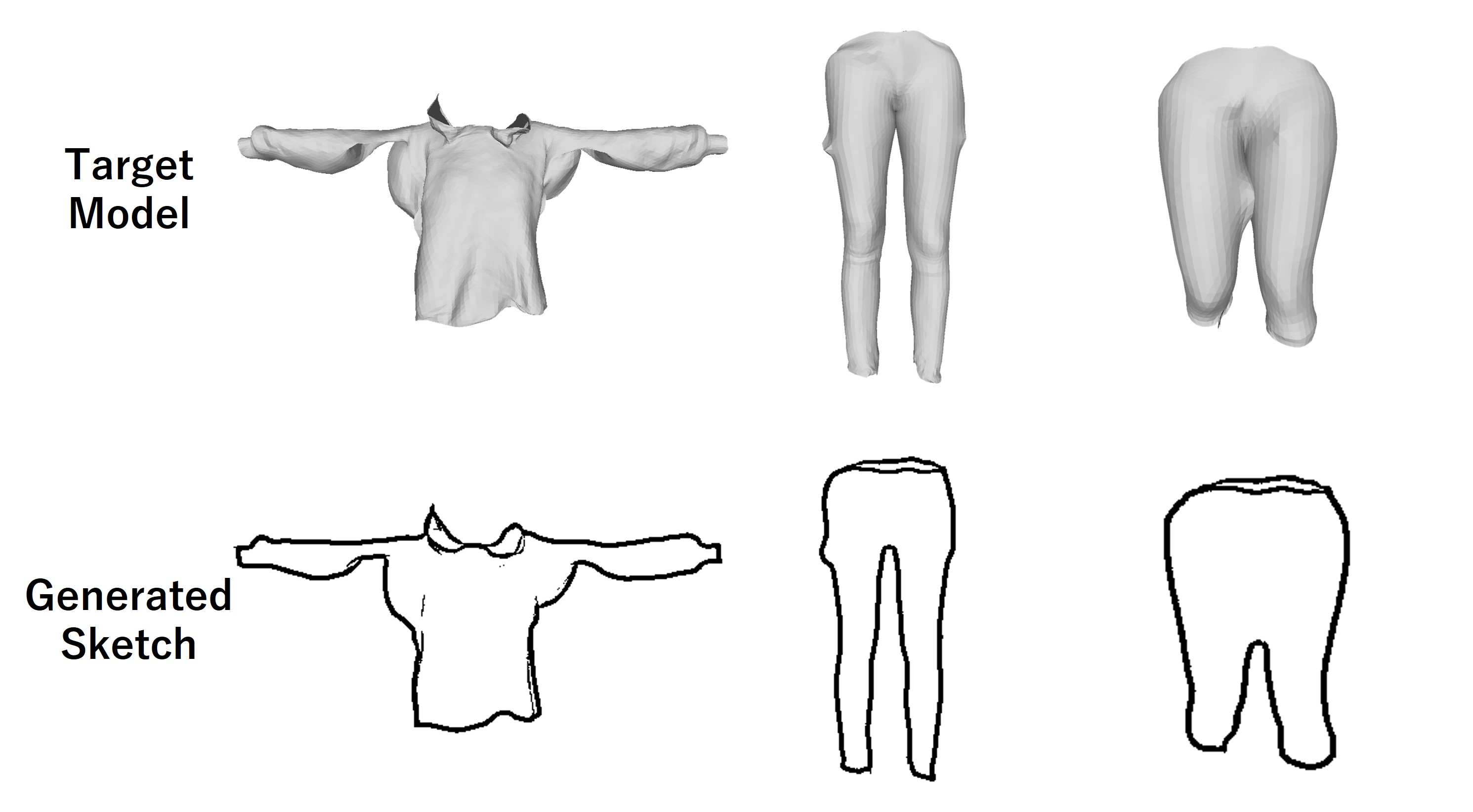}
    \caption{Generated sketch-mesh pair data. }
    \label{fig:example_sketch}
\end{figure}

The dataset used in this study does not provide paired sketch images. To obtain sketches, rendered images must first be generated for each model, and the generated rendered images are used to generate sketch images.
Considering the sparse features of the hand-drawn sketches, we extract the binarized contours of the target meshes from the rendered depth maps at different angles as sketch images. The depth map can remove unnecessary details to produce a higher-quality contour.
For each mesh $M_i$, we render depth maps from 36 angles.
 The mesh data paired with the sketch data obtained by this method is shown in Figure \ref{fig:example_sketch}.

\begin{figure}[h]
    \centering
    \includegraphics[width=0.9\linewidth]{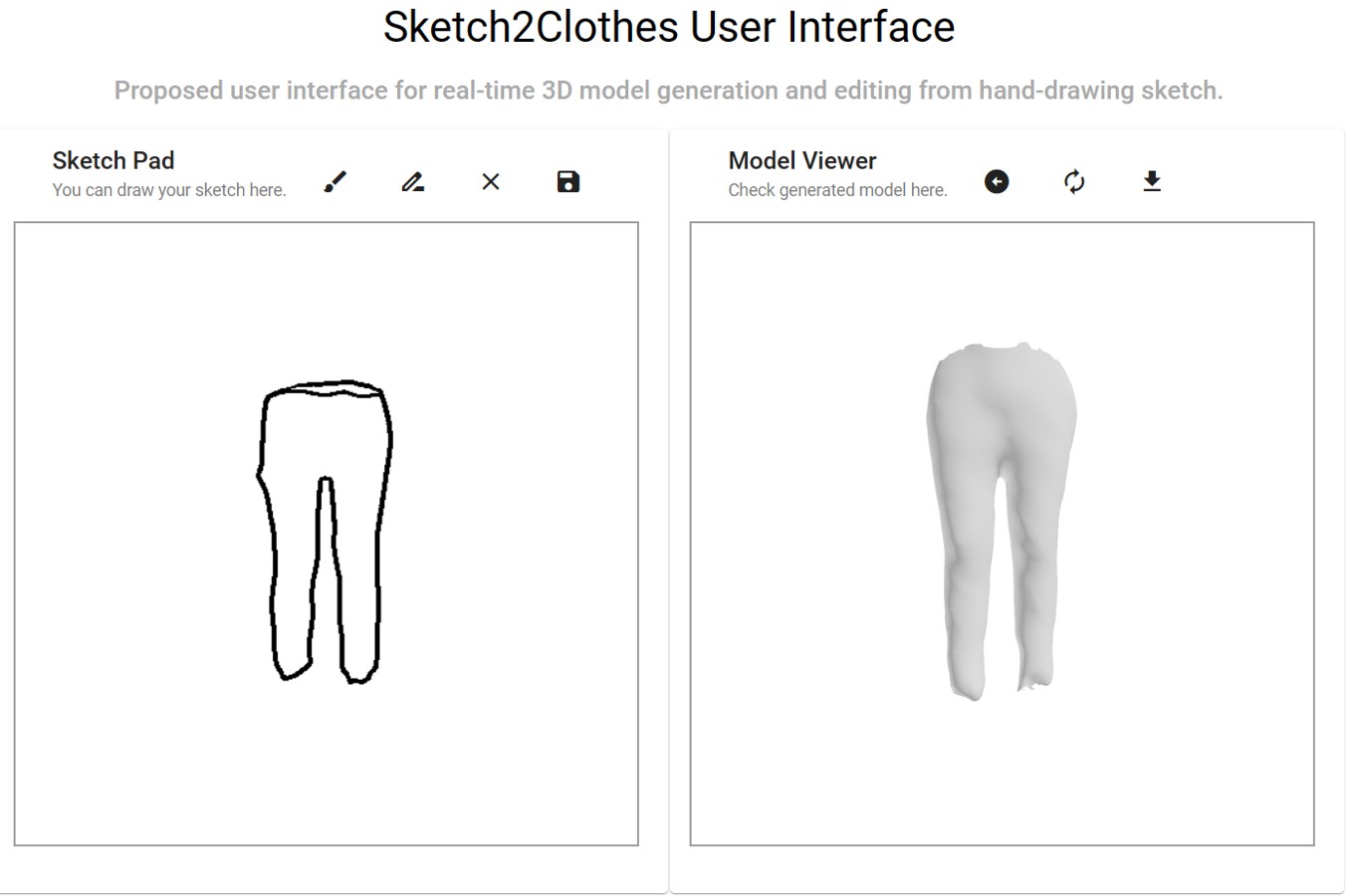}
    \caption{The proposed user interface of Sketch2Cloth.}
    \label{fig:user_interface}
\end{figure}

\subsection{User Interface}

The proposed system consists of 3D garment model generation from sketch input and a user interface for creating and editing sketches. The proposed user interface (UI) is shown in Figure \ref{fig:user_interface}, which consists of two areas. The left area is for creating and editing sketches, and the right area is for viewing generated objects.

All buttons are located at the top of the UI. In the "Sketch Pad" area, the buttons from left to right are ``Brush," ``Eraser," ``Clear," and ``Save." The user can use these functions to create a sketch, and when finished, press "Save" to start model generation from the created sketch. The ``Model Viewer" operation buttons are located at the top of the right side of the screen. From left to right are ``Capture," ``Reset," and ``Save Model," respectively. The generated object is displayed on the right side and can be zoomed in and out by scrolling the mouse and rotated by moving the mouse with the right mouse button pressed and held. The user can obtain a snapshot of the current viewpoint with the "Capture" button, and the system will generate a sketch image based on it. The generated image is placed in Sketch Pad and edited using the sketch creation function. After editing is complete, pressing "Save" one more time allows the system to optimize the current model based on the modified sketch and edit the model. The ``Reset" function dismisses the user's edits and returns to the initial generation result. Finally, the user can download and save the completed 3D model by pressing ``Save Model".

We implement the UI as a web application for wide applicability. The server side handles all the computational processing, so the system can be used without a burden on the user's terminal. We incorporate the Angular~\cite{angular} for the front end. For the server side, we choose Flask~\cite{flask} in order to connect the UI and the 3D model generation system. The average time cost was $1.13$ seconds from the user's creation of a sketch to the acquisition of the model.

\section{Implicit Surface Representation Learning}

\label{sec:net_learning}


In this study, we use the autoencoder structure~\cite{Park2019DeepSDFLC} for UDF training.
Let $E$ be an encoder that encodes the input sketch into a latent vector, $D$ be a decoder that generates UDF data, and $S$ be an input sketch. The input sketch $S$ is encoded into the latent code $z_c$ by the encoder $E$. The $Grid$ is the result of sampling a discrete normal Grid in the 3D space of \([-1, 1]^3\) where the 3D shape exists, denoted by $G$. The latent code $z_c$ is concatenated with $G$ to form the latent vector $z$, which is generated into a UDF by the decoder $D$. This process is shown in equation (\ref{equ:udf_generate}). The decoder predicts the unsigned distance $d_i^u$ for all the input coordinates $p_i \in G$.

\begin{equation}
    \label{equ:udf_generate}
    UDF = D(E(S), G)
\end{equation}

\subsection{Objective Function}

The learning objective of the network is to generate the best-fitting UDF for a given $S$. 
The optimization objective in learning is shown in equation (\ref{equ:learning_loss_udf}) by utilizing the $L1$ distance, where $df_{gt}^u$ is the ground truth.
To reduce the influence of sample coordinates scattered in the network, $df_{gt}^u$ and $D(z)$ are thresholded at the value of $\delta$ to remove discrete values described as $clamp$ function. 
Normalization is also performed on the input latent vector $z$, and optimization is also performed on the normalization results. The normalization loss is shown in equation (\ref{equ:learning_loss_reg}). In addition, a geometric regularization loss~\cite{long2022neuraludf} is adapted to improve the network learning as equation (\ref{equ:learning_loss_reg_geo}). The $\gamma$ and $\lambda$ refer to the normalization factor and the UDF values scalar, respectively. Based on the above, the loss function of the network is defined by the equation (\ref{equ:learning_loss}).

\begin{equation}
    \label{equ:learning_loss_udf}
    L(D(z), df_{gt}^u) = \mid clamp(D(z), \delta) - clamp(df_{gt}^u, \delta) \mid
\end{equation}

\begin{equation}
    \label{equ:learning_loss_reg_geo}
    L_{reg\_geo} = \frac{1}{N} \sum exp(-\gamma \cdot D(z))
\end{equation}

\begin{equation}
    \label{equ:learning_loss_reg}
    L(z_i) = \lambda \parallel z_i \parallel _2
\end{equation}

\begin{equation}
    \label{equ:learning_loss}
    L = L(D(z), df_{gt}^u) + L(z_i) + L_{reg\_geo}
\end{equation}

\begin{figure}[h]
    \centering
    \includegraphics[width=\linewidth]{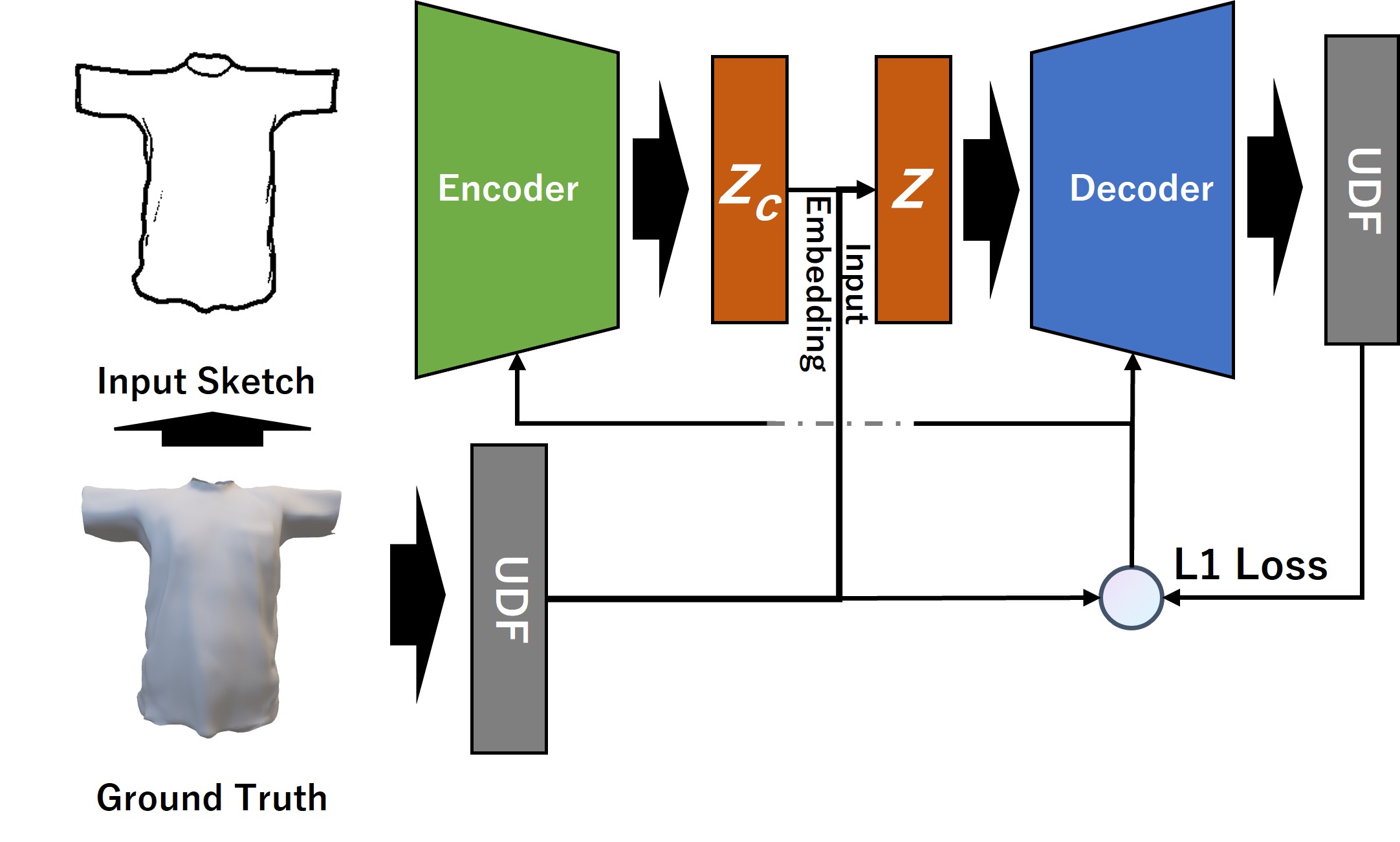}
    \caption{The learning flow of the network.}
    \label{fig:ae_learning_process}
\end{figure}



\begin{figure*}
    \centering
    \begin{subfigure}{0.6\linewidth}
      \centering
      \includegraphics[width=\textwidth]{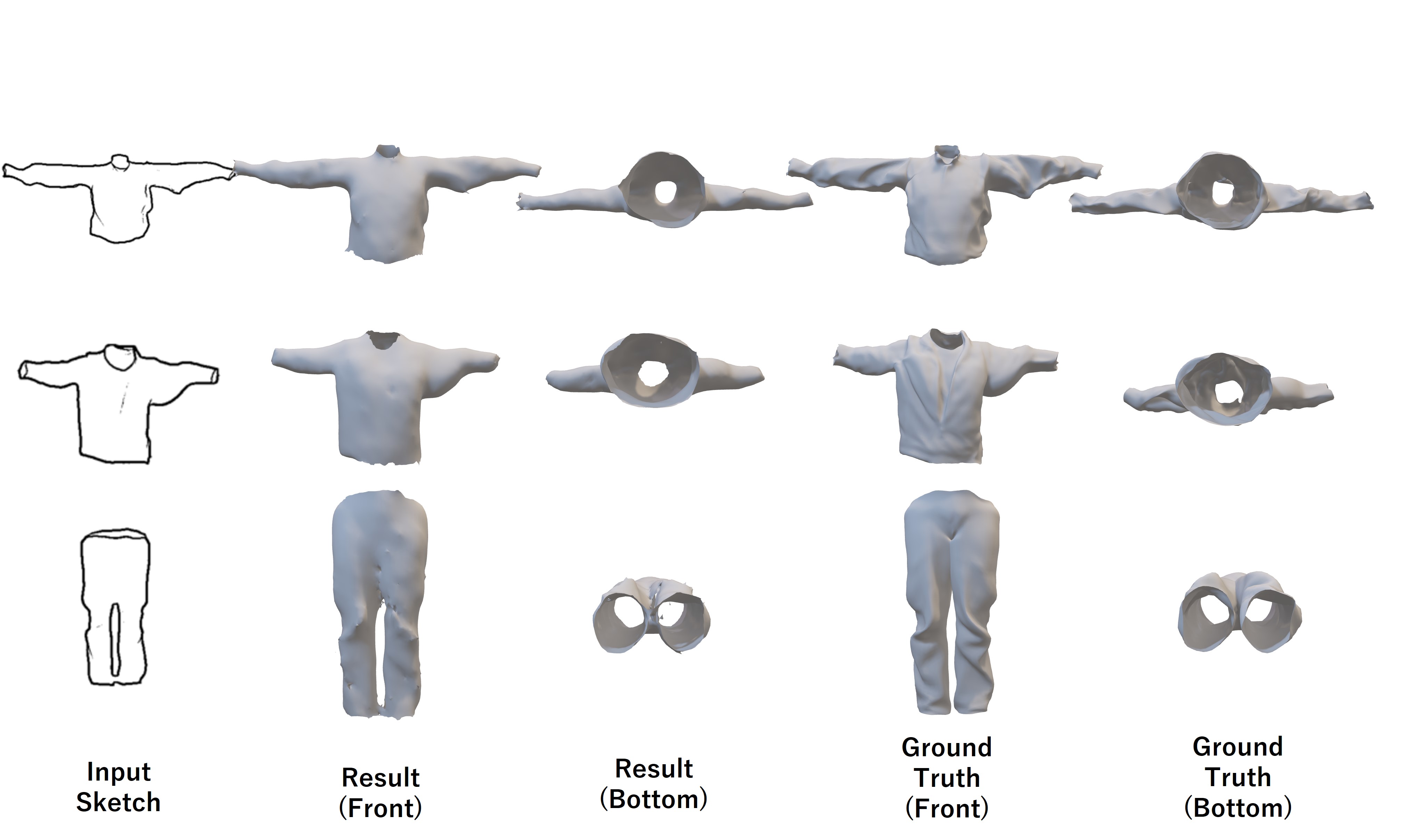}
        \caption{Multi-Garment}
        \label{fig:results_nonwatertight_syn}
    \end{subfigure}
    \begin{subfigure}{0.6\linewidth}
      \centering
        \includegraphics[width=\textwidth]{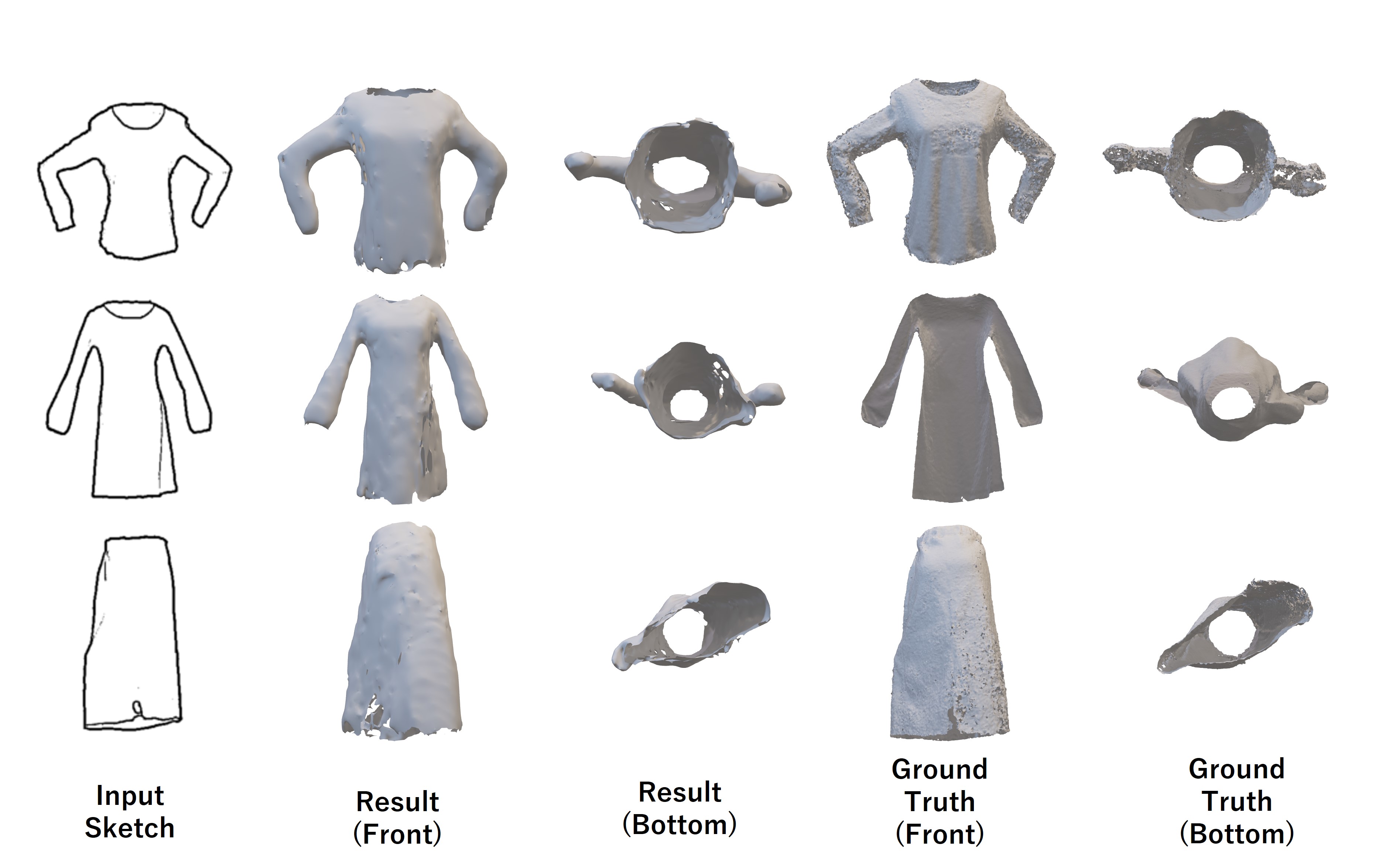}
        \caption{DeepFashion3D}
        \label{fig:results_nonwatertight_syn_df3d}
    \end{subfigure}
    \caption{Results of the garment model reconstructed from the rendering sketches on Multi-Garment (a) and DeepFashion3D datasets (b).}
\end{figure*}

\begin{figure*}[t]
    \centering
    \includegraphics[width=\linewidth]{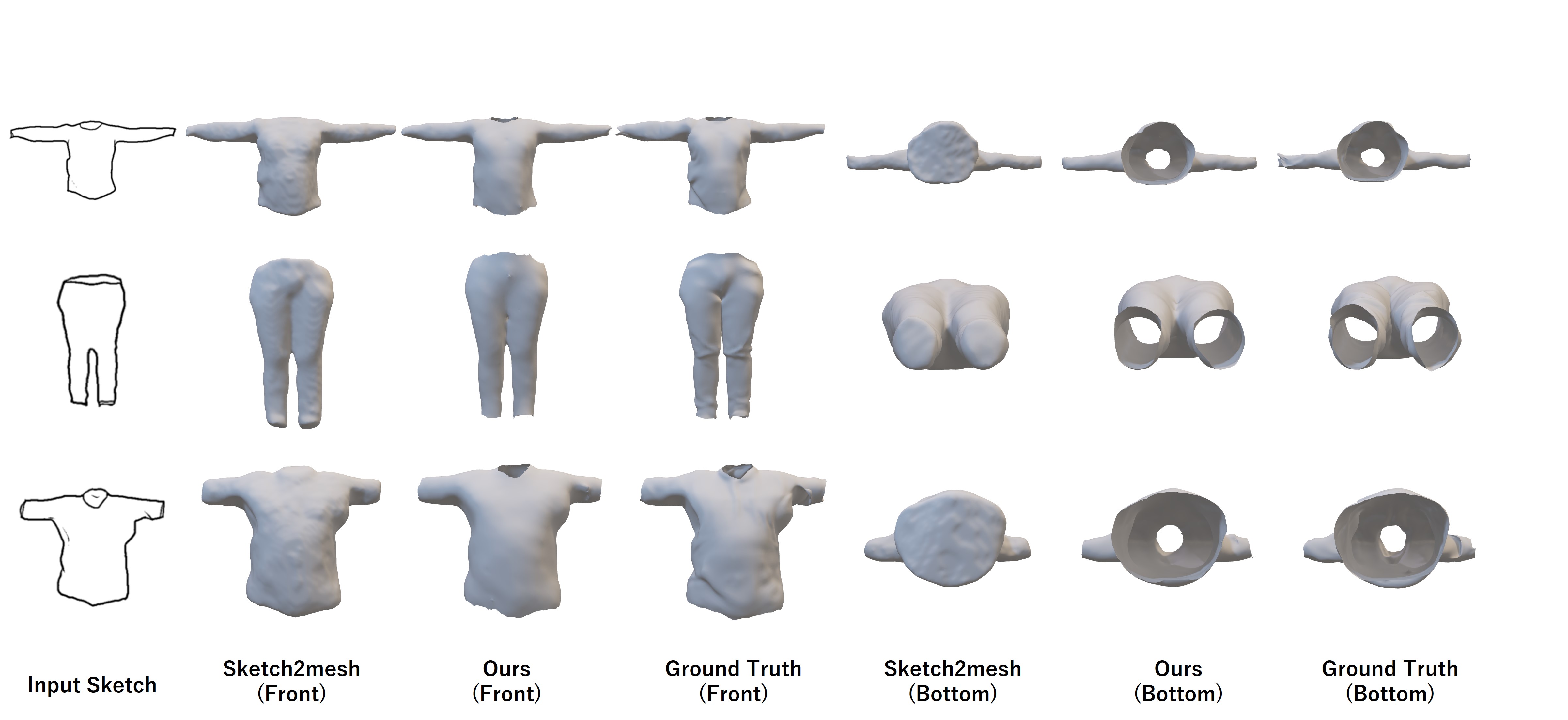}
    \caption{Comparison with state-of-the-art sketch-based model reconstruction method~\cite{Guillard2021Sketch2MeshRA}.}
    \label{fig:results_comparison}
\end{figure*}

\begin{figure}[h]
    \centering
    \includegraphics[width=\linewidth]{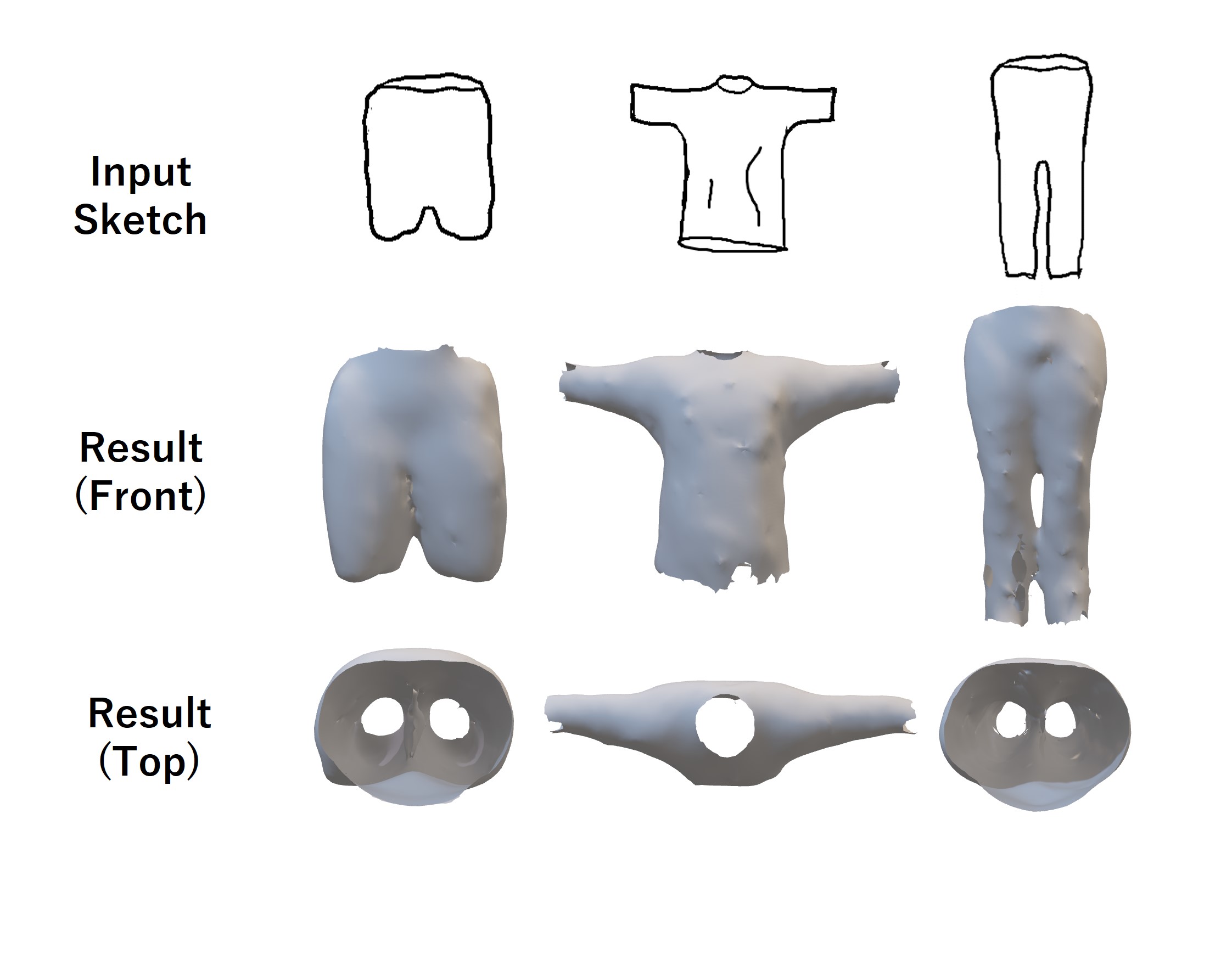}
    \caption{Results of the garment model reconstructed from the hand-drawn sketch.}
    \label{fig:results_nonwatertight_handdrawing}
\end{figure}

\subsection{Optimization for Generated Model}

In the optimization and editing of the generated results, this study refers to the method of Remelli et al.~\cite{MeshSDF}. The result and the chamfer distance of the sketch are minimized to optimize the latent vectors as decoder input. The chamfer distance is used as a limiting constraint for the optimization. In this study, the chamfer distance is implemented in 2D, where the 2D Chamfer Loss $L_{CD}$ is defined by the equation (\ref{equ:chamfer_loss}). where $S_p$ refers to the projected contour of the 3D mesh and $F_s$ represents the sketch region.

\begin{align}
    \label{equ:chamfer_loss}
    L_{CD} = \sum_{p_{p} \in S_{p}} min_{p_{s}|F_s(p_{s})=0} \parallel p_{p} - p_{s} \parallel ^2 \nonumber\\ +
    \sum_{p_{s}|F_s(p_{s})=0} min_{p_{p} \in S_{p}} \parallel p_{p} - p_{s} \parallel ^2
\end{align}


The optimization of the generated results essentially backpropagates the gradient values due to the loss between the generated results and the input, but since neither the UDF data nor the 3D model is differentiable, the optimization target is the latent vectors of the decoder inputs. Since the latent vectors are differentiable as part of the network learning process, optimization is possible. This optimization allows the output result of the network to be tuned. Specifically, if the generated 3D mesh is $M$ and the sampling of its surface vertices is $V$, the gradient $\frac{\partial L_{CD}}{\partial z}$ is calculated for the latent vector $z$ using the equation (\ref{equ:gradients_computation}). The calculated gradients are used to optimize the latent vectors through the backpropagate process.

\begin{equation}
    \label{equ:gradients_computation}
    \frac{\partial L_{CD}}{\partial z} = \sum_{v \in V} - \frac{\partial L_{CD}}{\partial v}\nabla D(v, z) \frac{\partial D}{\partial z}(v, z)
\end{equation}

The optimized latent vectors are then used to generate UDF data by the decoder and reconstruct the 3D model. 

\subsection{Network Training}

The deep learning part of the system was implemented in Pytorch~\cite{pytorch} and Pytorch3d~\cite{ravi2020pytorch3d}.
The Auto Encoder used in the UDF generation follows the implementation of Guillard et al.~\cite{MeshSDF}. The decoder consists of 9 perceptron layers, the number of hidden dimensions in the MLP layer is 512, and ReLU is used as the activation function. Fourier position encoding~\cite{Sitzmann2020ImplicitNR} is performed on the input 3D coordinates to reduce the loss of detailed information due to the activation function.
The encoder part, which encodes the sketch image into latent vectors, uses ResNet~\cite{He2016DeepRL}. In learning, the loss function is the sum of the $L1$ loss due to UDF (equation (\ref{equ:learning_loss_udf})) and the input normalization loss (equation (\ref{equ:learning_loss_reg})). Set the scatter removal factor $\delta$ to $0.1$ and the normalization strength $\lambda$ to $10^{-4}$. 
The training flow of the network is shown in Figure \ref{fig:ae_learning_process}.

The batch size was 16 and the number of epochs was $2000$. The optimizer used was Adam~\cite{Kingma2015AdamAM}, and the learning rate was set according to the number of epochs according to the equation (\ref{equ:learning_rate_schedule}). $lr_{init}$ is the initial learning rate, and the initial value is set to $0.5 \times 10^{-3}$. The $\omega$ is $0.5$ and ${\gamma}$ is 500. For the encoder, we set $\alpha = 1$, and for the decoder, $\alpha = 0.1$. The training was performed using two NVIDIA GeForce RTX 3090s. The training time per epoch was around $10$ seconds and about $5.5$ hours for all epochs. 

\begin{equation}
    \label{equ:learning_rate_schedule}
    learning\_rate = \alpha \times lr_{init} \times ( \omega ^ {\lfloor \frac{epoch}{\gamma} \rfloor} )
\end{equation}

\section{Evaluation}

Both the sketch generated by the rendering method and the user interface were used to evaluate this system. Note the user-drawn sketches are created by one of the authors. The effectiveness of the proposed method is also verified by numerical evaluation of the 3D garment models generated by the proposed method, as well as by comparison with state-of-the-art Sketch2Mesh approach~\cite{Guillard2021Sketch2MeshRA}).

\subsection{Model Generation from Sketches}

For the generation of the garment model by sketching, we use the garment data provided by Multi-Garment~\cite{bhatnagar2019mgn}. To verify the effectiveness of the system's generation, we use both sketched images generated from rendered images and hand-drawn sketches.

The results of generating 3D garment models from input sketches using this method are shown in Figure \ref{fig:results_nonwatertight_syn}. 
The system generates 3D non-watertight garment models according to the input sketches.
The output quality of the system did not change much from the results for hand-drawn sketches shown in Figure \ref{fig:results_nonwatertight_handdrawing}.
Figure \ref{fig:results_nonwatertight_syn_df3d} shows some results on DeepFashion3D dataset. we have to note that compare to Multi-Garment, the garment model provided by DeepFashion3D is not as good as Multi-Garment and contains a lot of holes, which affects our sketch generation and the learned UDF field.

\subsection{Comparison Evaluation}

As a state-of-the-art method for model generation by sketching, Sketch2Mesh~\cite{Guillard2021Sketch2MeshRA} is used for comparison. Because it is an SDF-based method, the non-watertight model used in this study is trained by water tightening it. The comparison results are shown in Figure \ref{fig:results_comparison}. The respective front and top views are shown. Existing SDF-based methods can reconstruct the model, but only as one continuous surface. Therefore, it is difficult to be useful for realistic reconstruction tasks. On the other hand, our method utilizes UDFs and can represent complex surfaces, such as non-watertight models. As a result of the numerical evaluation, the Chamfer Distance and Earth Mover's Distance of the generated model by the existing method and the proposed method are shown in Table \ref{tab:results_comparison}. It can be seen that the proposed method has a lower chamfer error and EMD error.

\begin{table}[h]
    \centering
    \caption{Chamfer Distance and Earth Mover's Distance of the generated model.}
    \begin{tabular}{c|c|c}
        \hline
        Method & CD(\text{$\times10^{-3}$}) & EMD(\text{$\times10^{-2}$}) \\
        \hline
        \hline
        sketch2mesh & 4.24 & 7.95 \\
        \hline
        ours & \textbf{3.51} & \textbf{7.14} \\
        \hline
    \end{tabular}
    \label{tab:results_comparison}
\end{table}

\subsection{Model Editing}

\begin{figure}[h]
    \centering
    \includegraphics[width=\linewidth]{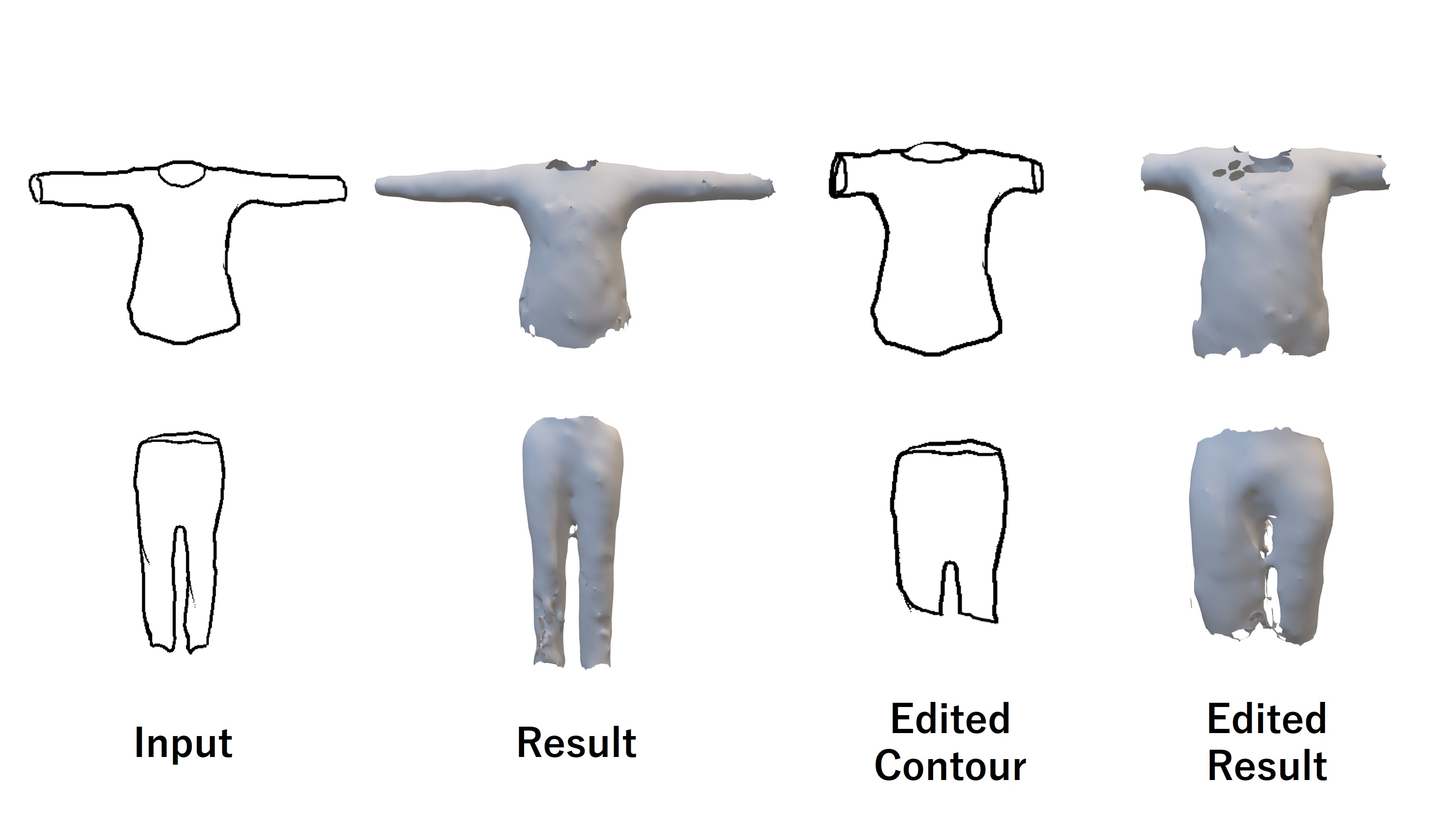}
    \caption{Results of editing reconstructed model}
    \label{fig:results_editing}
\end{figure}

As described in section \ref{sec:net_learning}, our system also allows users to edit the output models. As shown in Figure \ref{fig:results_editing}, users can edit the output 3D model via 2D sketches that are easy to manipulate, rather than directly editing the 3D model for areas where they are dissatisfied with the generated results.

\section{Conclusion}

This work proposed Sketch2Cloth for generating 3D clothing models from hand-drawn sketches, using implicit function learning (UDF) to generate non-watertight clothing models. Sketch2Cloth enables users who do not have expertise in 3D modeling to create 3D models of desired garments easily. In this work, UDF learning was applied to sketch-based generation to generate a 3D model of a garment.  We found that it may be difficult to represent the detailed surface feature, such as pockets and buttons. 

In this work, we used an autoencoder to learn the unsigned distance field. However, we found that the values and  gradients of UDF field could be difficult to learn due to the un-differentiability at the zero level set, which may result in the holes or wrong normal of reconstructed meshes. Recent works adopted RGB images for learning with more image features than the binary sketches~\cite{long2022neuraludf,Zhao2021LearningAU}. To solve this issue, We believe that an estimated normal map from the input sketch~\cite{he2021sketch} may learn the better UDF fields.


For the model generation, all garment parts (sleeves, buttons, pockets, etc.) are integrated, and it is expected that the actual design application will require the separation of each part. Although this study used UDF to successfully generate garment models from sketches, we believe that Sketch2Cloth can be extended to the generation of a car's exterior and interior design from sketch input, and interior scenes~\cite{Chibane2020NeuralUD}. It is also expected that the proposed system can be used to create complex scenes.



\bibliographystyle{IEEEtran}
\bibliography{IEEEabrv, refs}

\end{document}